# Network Control Systems RTAI framework – A Review

Deepika Bhatia[*], Urmila Shrawankar

*Department of Computer Science & Engineering, G.H.R.C.E. Nagpur*

*Abstract*--With the advancement in the automation industry, to perform complex remote operations is required. Advancements in the networking technology has led to the development of different architectures to implement control from a large distance. In various control applications of the modern industry, the agents, such as sensors, actuators, and controllers are basically geographically distributed. For effecient working of a control application, all of the agents have to exchange information through a communication media. At present, an increasing number of distributed control systems are based on platforms made up of conventional PCs running open-source real-time operating systems. Often, these systems needed to have networked devices supporting synchronized operations with respect to each node . A framework is studied that relies on standard software and protocol as RTAI, EtherCAT, RTnet and IEEE 1588. RTAI and its various protocols are studied in network control systems environment.

*Keywords*- Real time systems, RTAI, EtherCAT, Ethernet, RTOS

## I. INTRODUCTION

The correctness of real-time systems depends not only on the exact results of computation [1], but also on the time at which the results are produced. Real-time computing is that in which the system correctness not only depends on the correctness of logical result but also on the result delivery time. So the operating system should have features to support this critical requirement to be termed as a Real-time operating System(RTOS). The RTOS should have predictable behaviour to unpredictable external events. "A good RTOS is one that has a bounded behaviour under all system load scenario i.e. also under simultaneous interrupts and thread execution". Thus we can say that a true RTOS will be deterministic under all conditions. These operating systems occupy little space as compared to the General Operating systems which take hundreds of megabytes. Therefore, these systems are classified into two categories based on their timing constraints: hard real-time and soft real-time systems. In hard real-time systems, the violation of timing constraints of certain tasks should not be acceptable. For example, not executing a task before its deadline may lead to catastrophic consequences in some environments such as patient monitoring systems, nuclear plant control, and flight control etc. On the other hand, in soft real-time systems, although system performance is decreased when deadline is missed but it does not have serious damage on the system as these systems are fault tolerant. In real time systems throughput is not so much important, the responsiveness of the system matters. The transmission of continuous media of multimedia systems is one of the typical tasks in soft real-time system's environments. Hard real-time scheduling theory determines the scheduling ability of a system depending on the worst case execution times (WCET) of the tasks. The hard real-time scheduling methods cannot be used in soft real-time systems, because designing a soft real-time system using hard real-time scheduling theory often yields a system whose utilization is unacceptably low. The paper gives review to hard real time system RTAI and its framework, giving introduction to various protocols such as EtherCAT, RTnet and IEEE 1588 etc.

Basically the Ethernet is based on IEEE 802.3 that can support high-speed in LANs (local area networks). Unfortunately, this protocol has not supported real-time traffic earlier. Therefore, several researches of supporting the real-time traffics on Ethernet have been proposed. Such approaches either implemented or proposed the real-time guarantees based on switch. Some of these approaches also supported both hard and soft real-time guarantees. EDF scheduling algorithm is studied earlier to verify the scheduling feasibly and also to dispatch the coming packets. There are two different approaches to provide real-time performance with Linux: Improving the Linux kernel pre-emption and adding a new software layer beneath Linux kernel with full control of interrupts and processor key features. RTAI is real time application interface which implements the above mentioned Linux features. RTAI has extended its API to allow remote (other host) procedure call RPC. New API functions have been added with the following syntax: replace the first two letters of the function name (for example: given the rt_mbx_send(), the new function RT_mbx_send() has been added); and the new function has two new parameters: node and port. This feature do not comply with any communication standard. The EtherCAT technology is given which overcomes the system limitations of other Ethernet solutions. RTnet is studied which is a purely software-based framework for exchanging arbitrary data under hard real-time constraints.

The rest of the paper is organized as follows:- In section II related work from previous papers is presented. Also various protocols have been studied. Advantages and limitations of various techniques are discussed. Section III gives the conclusion drawn from the review paper. Section IV gives the problems and future directions that can help to explore the related issues.

## II. RELATED WORK

Hard real-time communication over Ethernet has been a popular research topic for more than a decade. Various algorithms and technologies have been discussed and evaluated.

T. Chiueh et al., [2] proposed a token-based approach which is further studied by F. Hanssen et al., [3]. In token-based real-time networks, nodes transmit data when they get the token. Token passing needs proper mechanism for sending data over the transmission media so that no token could get lost or gets duplicated during the sending process. Thus the transmission media has to bear the overhead. Another approach to guarantee real-time constraints over Ethernet is to restrict the amount of traffic which is transmitted by each network node in a certain period of time. J. Loeser et al., [4] provides a model for traffic shaping used in switched Ethernets networks.

Lorenzo Dozio et al. [5] studied issues related to the design and implementation of high performance distributed control systems with reference to the use of RTAI services and mechanism. He studied two kinds of issues given below:-
- "Can Ethernet be Real Time?"
- Should implementation of DRTOSes (distributed real time operating systems) be based on application specific hardware/software or on off the shelf technologies?

Assumption has been made that to meet profiling specifications some form of active structural control for compensating the machine frame compliance and vibrations is required, so the sampling rate should be as high as 10 KHz. If designed application is run standalone on a GPCPU(general purpose central processing units) and its execution timing checked, they found latencies in the range of few microseconds. He discussed the real time application interface (RTAI) , services provided by RTAI and RTAI-Lab in detail.

*A. RTAI overview*

The Real Time Application Interface (RTAI) project began at the "Dipartimento di Ingegneria Aerospaziale del Politecnico di Milano" in 1996/97 which provides a hard real time extension to Linux. It is a cost effective tool to support a activities related to advanced active controls for generic aeroservoelastic systems and real time simulation. RTAI is integrated into Linux through a text file containing a set of changes to its kernel source code, and a series of add on programs expanding Linux to hard real time by installing a generic Real T ime Hardware A bstraction L ayer ( RTHAL). Figure 1 shows simplified diagram of RTAI and Linux kernel architecture [10]. RTHAL performs three primary functions:
- It collects all the pointers to the time critical kernel internal data and functions into a single structure, to allow the easy trapping of all the kernel functionalities that are important for real time applications, so that they can be dynamically substituted by RTAI when hard real time is needed.
- "Reworks the related Linux functions, data structures and macros to make it possible to use them to initialize RTHAL pointers for normal Linux operations."
- "Changes Linux to use what pointed in RTHAL for its operations."

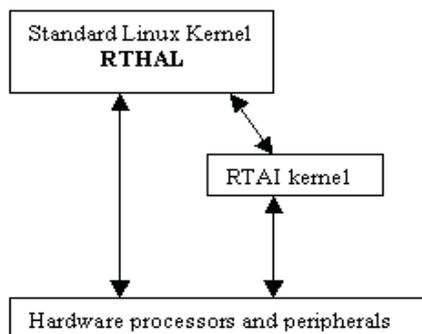

Figure 1 Simplified RTAI and Linux Kernel Architecture

*B. RTAI Services:*

RTAI does provide the required scheduler along with a different types of services. The RTAI scheduler is fully preemptible and can schedule directly from within interrupt handlers so that Linux could not delay any RTAI hard real time activity, as the delay could be fatal. RTAI schedulers gives the following scheduling policies:
- First In First Out (FIFO) fully preemptable service for scheduling. It helps to meet periodic tasks deadlines with statically assigned priorities .
- Round Robin (RR) is like FIFO but a time quantum is alloted, after which the CPU is given to any equal priority task waiting on the ready list, for some time.
- RTAI provides its own small (20KBs) and effective real time middleware layer, called net_rpc. It integrates distributed and local applications by the addition of a port identifier in front of any RTAI function call.
- Early Deadline First (EDF), scheduling for dynamically assigning priorities in order to meet deadlines of periodic tasks.

*C. RTAI-Lab*

It provides a framework to design, build, run and monitor any suite of RTAI whether locally or in distributed way. Also Syndex an automatic code generator is integrated in RTAI-lab. It can optimize distribution of controllers on available CPUs. It also takes the communication delays into consideration.

Won- jong Kim et al. [7] suggested to develop networked control systems (NCSs) to implement distributed control from a distance. NCS consists of multiple nodes communicating with each other over the communication networks. A real-time operating environment is required for the implementation of an NCS for handling the timings of various events in the communications between these nodes. Different factors, such as time resolution and capability of multi-threading and periodic tasks, affect the selection of an appropriate real-time computing environment.

The presented software architecture is based on the UDP (user datagram protocol). The main focus is to develop a real-time operating environment for the closed-loop real-time control over distributed network. An NCS with closed-loop control of a maglev test bed is implemented on an LAN in real-time operating environment. It also incorporates a novel predictor-based algorithm to stabilize the system in case of successive network delays and data-packet losses.

Jan Kiszka et al., [6] suggested that real-time Ethernet is required to replace traditional fieldbuses. Earlier approaches for example FTT-Ethernet, RT-EP, are the combination of switches and traffic shapers. These approaches have various transport and application protocols which are generally not compatible with each other. Also, there are other transport media beyond Ethernet 100Base-T approaching the real-time domain: Gigabit Ethernet, wireless media as IEEE 802.11 or Bluetooth, and FireWire. Rtnet is given which is hardware independent and flexible real-time communication platform, and provides deterministic networking. RTnet is a purely software-based framework for exchanging arbitrary data under hard real-time constraints. It provides scalability and extensibility according to the application. It also enables the integration of various other communication media besides Ethernet.

*A. Base Services*

RTnet contains a set of central services which are required for most scenario. For example:-

- **Packet Management**
  Packets are passed through the stack in the context of the sending task.
- **UDP/IP Implementation**
  In it the dynamic Address Resolution Protocol (ARP) was converted into a static mechanism which is executed during the set-up. Also, the routing process was simplified. The output routing tables were optimized for the limited amount of data.

*B. FireWire Integration*

FireWire, also known as IEEE 1394, is a high performance serial bus for connecting different types of devices. Initially it was targeted for consumer-electronic applications, such as video transmission of high speed. The RTnet mechanism [6] for real-time packet management is applied to the FireWire stack as well. NIC driver and high-level applications are both producers and consumers of packets. RTOS semaphore is used for the synchronization between server and tasklet queue. The server runs at a higher priority than application tasks. The connection between FireWire stack and RTnet core is implemented through Ethernet emulation. By the use of Ethernet emulation, FireWire functions same as other real-time Ethernet devices in RTnet.

Ianluca Cena et al., [8] proposed EtherCAT a solution based on real-time Ethernet (RTE). It uses a technique to exchange data with I/O devices which resembles closely the summation frame adopted in earlier fieldbus networks, such as Interbus-S. In it, the communication efficiency achieved by the designed module is very high It achieves synchronization by distributed clock (DC) mechanism. It reduces the implementation costs.

*A. Distributed clock mechanism*

In the paper DC-enabled EtherCAT slaves are provided with an internal clock whose nominal period is 10ns. The current time value is held in an internal register of 64-bit, with a granularity of 1ns. It represents the time elapsed from the January 1st, 2000. With EtherCAT, the data exchange is completely hardware based on "mother" and "daughter" clocks. Each clock can simply and accurately determine the other clocks run-time offset because the communication utilizes a logical and full-duplex Ethernet physical ring structure. The distributed clocks are adjusted based on this value, which means that a very precise network-wide timebase with a jitter of significantly less then 1 microsecond is available. However, high-resolution distributed clocks are not only used for synchronization, but can also provide accurate information about the local timing of the data acquisition.

*B. Performance of EtherCAT*

The extremely high performance of the EtherCAT technology enables control concepts that could not be realized with classic fieldbus systems. For example, the Ethernet system can not only deal with velocity control, but also with the current control of distributed drives. The tremendous bandwidth enables status information to be transferred with each data item. With EtherCAT, a communication technology is available that matches the superior computing capacity of modern Industrial PCs. The bus system is no longer the "bottleneck" of the control concept. Distributed input and outputs are recorded faster than is possible with most local I/O interfaces.

Gianluca Cena et al.,[9] suggested an architecture for a real-time distributed system based on RTAI, RTnet and the PTP(precision time protocol) protocol. The main advantage of such an architecture is that, it can be ported on different H/W and S/W platforms. A prototype has been implemented and its performance is verified through different experimental measurements and methods. The industrial Ethernet solutions tackle synchronization with the help of various defined protocols. For general purpose operating systems IEEE 1588 precision time protocol (PTP) is a better solution. Packet based networks like Ethernet are inherently non-deterministic. In order to gain determinism, distributed real-time applications need to be decoupled by a deterministic abstraction layer. If all nodes are equipped with highly synchronized real-time clocks, determinism can be achieved. The objectives of the IEEE 1588 standard are :-

- Highly synchronized real-time clocks in components of a networked distributed measurement and control system
- Intended for relatively localized systems typical of industrial automation and test and measurement environments.
- Applicable to local area networks supporting multicast communications (including but not limited to Ethernet)
- Supports heterogeneous systems of clocks with varying precision, resolution and stability
- Minimal resource requirements on networks and host components.

Through PTP, multiple devices are automatically synchronized with the most accurate clock found in a packet-based network, typically Ethernet. The RTS (real time stack) protocol stack automatically determines the most accurate clock, otherwise known as the Grand Master Clock. During operation and after initial synchronization, the PTP real-time clocks are constantly adjusted by exchanging timing messages. The RTS implementation uses statistical techniques to further reduce residual fluctuations. Because the RTS IEEE 1588 protocol stack supports the PTP hot-pluggable functionality requirement, devices may join or leave the network at any time. Two important assumptions have to be done for the PTP mechanism do perform in a correct manner:-

- The time between message exchanges should be small.
- The time required for a message to travel from the master to the slave should be equal to the time it takes to go from the slave to the master.

S/W based PTP protocol is implemented in the paper. Its architecture is mostly inexpensive. With RTAI Linux extensions, a complete distributed real-time system is designed. RTnet framework in to support hard real-time communications over Ethernet is adopted in the paper. The use of both RTAI and RTnet achieves good accuracy, even in case of heavy I/O load. He also studied that the implementation does not require any specific H/W, no changes are required to NIC drivers. Also the application is not affected by interferences from lower-priority tasks, even under high system loads

### III. CONCLUSION

Standard Linux is not able to assure real-time guarantees for applications. Several extensions, which permit the use of Linux in real-time systems, have been proposed earlier. The paper focuses on RTAI, which runs standard Linux as low-priority task within a separate real-time kernel. The open source protocol stack RTnet provides deterministic communication over non-deterministic media like Ethernet. RTAI and RTnet are able to meet deadlines under high CPU loads, memory interactions invalidate real-time constraints when more data than physical CPU caches can handle is transferred at once. RTnet is an open source hard real-time network protocol stack for RTAI which provides real-time communication over Ethernet. RTnet introduces an additional protocol layer called RTmac to avoid collisions on the Ethernet. EtherCAT protocol is studied for use in factory automation environment. IEEE 1588 precision time protocol is studied that provides a method for the nodes to choose master-slave mechanism for clock synchronization.

### IV. PROBLEMS AND FUTURE DIRECTIONS

Further integration of FireWire, new media like Gigabit Ethernet, and interoperation with additional middlewares can be studied upon. Based on the connection to RTnet via Ethernet emulation, the adoption of FireWire's transaction modes and clock synchronisation for RTnet services can be addressed by researchers in future. CANopen over RTnet can be analysed and can lead to the implementation of an extended CANopen stack. We can plan to study the interferences between the real-time tasks used for PTP synchronization and the control application task, by varying the network load and by using a different a real-time scheduler such as, e.g, Xenomai.

### REFERENCES

[1] Sedigheh Asiaban et al. *A Real-Time Scheduling Algorithm for Soft Periodic Tasks* International Journal of Digital Content Technology and its Applications Volume 3, Number 4, December 2009
[2] T. Chiueh and C. Venkatramani, "*Supporting Real-time Traffic on Ethernet*", Proc. of IEEE Real-time Systems Symposium, Dec. 1994, pp. 282-286.
[3] F. Hanssen, P.G. Jansen, H. Scholten, S. Mullender, "*RTnet: a distributed real-time protocol for broadcast-capable networks*", Joint Int. Conf. on Autonomic and Autonomous Systems and Int. Conf. on Networking and Services, 2005. ICAS-ICNS 2005.
[4] J. Loeser and H. Haertig. "*Low-latency hard real-time communication over switched ethernet*". 16th Euromicro Conf. on Real-Time Systems, Catania, Sicily, July 2004, pp. 13- 22.
[5] Lorenzo Dozio et al. "Real Time Distributed Control Systems using RTAI",Proceedings of the Sixth IEEE International Symposium on Object-Oriented Real-Time Distributed Computing (ISORC'03) 0-7695-1928-8/03 2003 IEEE
[6] Jan Kiszka, Bernardo Wagner et al. "RTnet – A Flexible Hard Real-Time Networking Framework" 0-7803-9402-X/05/ 2005 IEEE
[7] Won-jong Kim,,"Real-Time Operating Environment for Networked Control Systems", IEEE, VOL. 3, NO. 3, JULY 2006
[8] Ianluca Cena et al. ,"On the Accuracy of the Distributed Clock Mechanism in EtherCAT", IEIIT, 2010
[9] Gianluca Cena et al,"A Software Implementation of IEEE 1588 on RTAI/RTnet Platforms" . 978-1-4244-6850-8/10/ 2010 IEEE
[10] Hao Cai, *A Predictable and IO Bandwidth Reservation Task Scheduler* ,Nan Kai University, pdf, 2005